\documentclass[12pt]{article}
\usepackage[czech]{babel}
\usepackage{graphicx}

\title {\bf An implication in orthologic}
\author {I. Chajda, R. Hala\v {s} \footnote {The paper was prepared under the support of 
Czech Government Council No. 314/98:153100011.}} 

\date{}

\begin{document}

\maketitle

{\bf Abstract.} We involve a certain propositional logic based on an ortholattice.
We characterize the implication reduct of such a logic and show that its algebraic 
counterpart is the so-called orthosemilattice. Properties of congruences and 
congruence kernels of these algebras are described.


{\bf MS classification:} 06C15, 03G25, 08A30.

\vskip8pt

By an {\bf ortholattice} is meant an algebra ${\cal L}=(L;\vee ,\wedge ,^\bot ,0, 1)$ of type
$(2,2,1,0,0)$ such that $(L;\vee ,\wedge )$ is a lattice with the least element $0$ and 
the greatest one $1$ and $^\bot $ denotes a complementation which is involutory, i.e.
$x^{\bot \bot }=x$ for each $x\in L$ and $x\leq y$ in $L$ implies $y^\bot \leq x^\bot $
(which is equivalent to De Morgan laws: $(x\vee y)^\bot =x^\bot \wedge y^\bot $ and
$(x\wedge  y)^\bot =x^\bot \vee~y^\bot $). Of course, every Boolean algebra and every
orthomodular lattice are ortholattices. However, a Boolean algebra serves as algebraic
counterpart of classical propositional logic where $\vee $ or $\wedge $ stand for disjunction 
or conjuction, respectively, and the complement $x^\prime $ of $x$ as a negation. Then the 
logical connective implication can be derived by
$$x\Rightarrow y \ = \ x^\prime \vee y.$$
On the other hand, an orthomodular lattice can analogously serve as an algebraic counterpart
of the so-called logic of quantum mechanics, shortly the so-called orthomodular logic, see [3].
In such a logic, the connective implication is expressed by means of $\vee ,\wedge $ and
complementation as follows:
$$x\Rightarrow y \ = \ (x^\bot \wedge y^\bot)\vee y.$$
Unfortunately, in ortholattices the analogy does not work. If we consider an ortholattice
visualized in Fig.1, then for $x\Rightarrow y:=x^\bot \vee y$ we have

\noindent
\hspace*{4.4cm} $a\Rightarrow b=1$ and $b\Rightarrow a=1$

\noindent
which contradicts to the accepted logical rules. 

\begin{figure}[h]
\begin{center}
\setlength{\unitlength}{1mm}
\linethickness{0.4pt}
\begin{picture}(40,40)(0,0)
\put(25,5){\line(3,4){6}}
\put(25,5){\line(-3,4){6}}
\put(25,5){\circle*{2}}
\put(31,13){\circle*{2}}
\put(19,13){\circle*{2}}
\put(24,0){$0$}
\put(35,12){$b^{\bot}$}
\put(12,12){$a$}
\put(31,13){\line(0,1){10}}
\put(19,13){\line(0,1){10}}
\put(31,23){\circle*{2}}
\put(19,23){\circle*{2}}
\put(35,23){$a^{\bot}$}
\put(12,23){$b$}
\put(19,23){\line(3,4){6}}
\put(31,23){\line(-3,4){6}}
\put(25,31){\circle*{2}}
\put(24,35){$1$}
\end{picture}
\title{Fig.1}
\end{center}
\end{figure}

\vskip8pt


Hence, we improve the object of our considerations as follows:

\vskip8pt

{\bf Definition 1.}
An ortholattice ${\cal L}=(L;\vee ,\wedge ,^\bot ,0, 1)$ is called a {\bf strong ortholattice} if for each $p\in L$ the interval
$[p,1]$ is also an ortholattice with respect to induced order, i.e. 
$([p,1];\vee ,\wedge , \ ^\bot _p,p, 1)$ is an ortholattice where for $a,b\in [p,1]$ the
operations $a\vee b, a\wedge b$ coincide with those of ${\cal L}$ and there exists an 
orthocomplement $a^{\bot }_{p}$ in $[p,1]$ for each $a\in [p,1].$

\vskip8pt

{\bf Example.}
A strong ortholattice which is neither modular (since 
\newline 
$\{0, e, d,b^\bot ,1 \}$ is a 
sublattice isomorphic to $N_5$) nor orthomodular (since $a\leq c^\bot $ but 
$a\vee (a^\bot \wedge c^\bot )=a\not=c^\bot $) is depicted in Fig. 2. 
 
\begin{figure}[h]
\begin{center}
\setlength{\unitlength}{1mm}
\linethickness{0.4pt}
\begin{picture}(60,40)(0,0)
\put(30,5){\line(1,1){7}}
\put(30,5){\line(-1,1){7}}
\put(30,5){\line(0,1){8}}
\put(30,5){\line(3,1){24}}
\put(30,5){\line(-3,1){24}}
\put(30,5){\circle*{2}}
\put(38,13){\circle*{2}}
\put(30,13){\circle*{2}}
\put(22,13){\circle*{2}}
\put(53,13){\circle*{2}}
\put(7,13){\circle*{2}}
\put(28,0){$0$}
\put(7,13){\line(0,1){8}}
\put(7,21){\circle*{2}}
\put(53,13){\line(0,1){8}}
\put(53,21){\circle*{2}}
\put(45,21){\circle*{2}}
\put(30,21){\circle*{2}}
\put(15,21){\circle*{2}}
\put(7,13){\line(1,1){8}}
\put(53,13){\line(-1,1){8}}
\put(22,13){\line(-2,1){15}}
\put(22,13){\line(1,1){8}}
\put(38,13){\line(-1,1){8}}
\put(38,13){\line(2,1){15}}
\put(30,13){\line(2,1){15}}
\put(30,13){\line(-2,1){15}}
\put(30,29){\circle*{2}}
\put(30,21){\line(0,1){8}}
\put(7,21){\line(3,1){24}}
\put(53,21){\line(-3,1){24}}
\put(15,21){\line(2,1){15}}
\put(45,21){\line(-2,1){15}}
\put(28,32){$1$}
\put(40,11){$d$}
\put(18,11){$a$}
\put(3,11){$e$}
\put(26,11){$b$}
\put(56,11){$c$}
\put(1,19){$c^{\bot}$}
\put(18,19){$d^{\bot}$}
\put(24,19){$b^{\bot}$}
\put(38,19){$a^{\bot}$}
\put(56,19){$e^{\bot}$}
\end{picture}
\title{Fig.2}
\end{center}
\end{figure}

For our purposes, a weaker structure is convenient, i.e. we will consider only order-filters
in a strong ortholattice which will be called orthosemilattice, precisely:

\vskip8pt

{\bf Definition 2.}
Let ${\cal S}=(S,\vee )$ be a semilattice with the greatest element 1 where for each $p\in S$
the interval $[p,1]$ is an ortholattice with respect to induced order; denote by $a^\bot _p$
the orthocomplement of $a\in [p,1]$ in $[p,1]$ and for $([p,1];\vee ,\wedge , \ ^\bot _p,p , 1)$
we have
$$a\wedge b=(a^\bot _{p}\vee b^\bot _{p})^\bot _{p},$$ 
where $\vee $ coincides with that of ${\cal S}.$ Then ${\cal S}$ is called an 
{\bf orthosemilattice.}

\vskip8pt

As it was already mentioned, each order-filter in a strong ortholattice is an orthosemilattice.
Every orthosemilattice is a set-theoretical union of strong ortholattices where the operations
$\vee $ and $\wedge $ coincide on the overlapping parts.

\vskip8pt

{\bf Theorem 1.}
{\it Let\/} ${\cal S}=(S;\vee )$ {\it be an orthosemilattice. Define the operation\/}
"$\bullet$" {\it as follows:\/}
$$x\bullet y:=(x\vee y)^\bot _{y}.$$
{\it Then\/}

\noindent
\hspace*{0.5cm} (a) $a\bullet 1=1, a\bullet a=1, 1\bullet a=a$

\noindent
\hspace*{0.5cm} (b) $(a\bullet b)\bullet b=(b\bullet a)\bullet a$

\noindent
\hspace*{0.5cm} (c) $(((a\bullet b)\bullet b)\bullet p)\bullet (a\bullet p)=1$ 

\noindent
\hspace*{0.5cm} (d) $(((a\bullet p)\bullet p)\bullet p)\bullet ((a\bullet p)\bullet p)=(a\bullet p)\bullet p.$

\vskip8pt

Proof.
{\bf (a)} Clearly $a\bullet 1=(a\vee 1)^\bot _1=1$

\noindent
\hspace*{3.9cm} $a\bullet a=(a\vee a)^\bot _a=a^\bot _a=1$

\noindent
\hspace*{3.9cm} $1\bullet a=(1\vee a)^\bot _a=1^\bot _a=a.$

\noindent
{\bf (b)} Since $a\vee b\geq b$ then $a\vee b\in [b,1]$, also $(a\vee b)^\bot _b\in [b,1]$
and hence $(a\vee b)^\bot _b\geq b.$ Then 
$(a\bullet b)\bullet b=((a\vee b)^\bot _b \vee b)^\bot _b =((a\vee b)^\bot _b)^\bot _b=a\vee b.$
Analogously, $(b\bullet a)\bullet a=b\vee a=a\vee b=(a\bullet b)\bullet b.$

\noindent
{\bf (c)} Since $a\vee b=(a\bullet b)\bullet b$, (c) can be rewritten as 
$((a\vee b)\bullet p)\bullet (a\bullet p)=1.$
It is easily seen that (c) is equivalent to 
$$a\leq b \ \Rightarrow  \ b\bullet p\leq a\bullet p.$$

\noindent
Suppose $a\leq b.$ Then $p\leq a\vee p\leq b\vee p.$ Since the orthocomplementation
in $[p,1]$ converses the order, we obtain
$a\bullet p=(a\vee p)^\bot_p\geq (b\vee p)^\bot_p=b\bullet p.$

\noindent
{\bf (d)} Similarly, (d) is equivalent to the condition 

\noindent
(d') \hspace*{3cm} $p\leq a \ \Rightarrow \ ((a\bullet p)\bullet a)\bullet a=1.$

Indeed, let (d) hold and $p\leq a$. Then $(a\bullet p)\bullet a=[(a\vee p)\bullet p]\bullet (a\vee p)=(((a\bullet p)\bullet p)\bullet p)\bullet ((a\bullet p)\bullet p)=a$, whence 
$$((a\bullet p)\bullet a)\bullet a=1.$$
Conversely, assume that $p\leq a$ and $((a\bullet p)\bullet a)\bullet a=1$. Then 
$(a\bullet p)\bullet a\leq a$ and since $a\leq (a\bullet p)\bullet a$, we have $(a\bullet p)\bullet a=a$.
Finally, replacing $a$ by $a\vee p$ in the previous equality, we obtain
$$[(a\vee p)\bullet p]\bullet (a\vee p)=a\vee p,$$
hence (d) holds.

If $p\leq a$, i.e. $a\in [p,1]$, then
$((a\bullet p)\bullet a)\bullet a=(a\bullet p)\vee a=(a\vee p)_p^\bot \vee a=a_p^\bot \vee a=1.$
\hfill $\rule{2mm}{2mm}$

\vskip8pt

{\bf Definition 3.} An algebra ${\cal A}=(A; \bullet, 1)$ of type $(2,0)$ satisfying 
the identities (a),(b),(c),(d) of Theorem 1 will be called an {\bf implication orthoalgebra.}

\vskip8pt

{\bf Remark.} The name implication orthoalgebra is motivated by the fact that the operation
"$\bullet $" can be considered as the logical connective implication. For the sake of brevity,
we shall write $x\bullet y$ instead of $x\Rightarrow y,$ analogously as in [1] where
this operation stands for the implication in a classical logic.

\vskip8pt

{\bf Theorem 2.}
{\it Let\/} ${\cal A}=(A;\bullet ,1)$ {\it be an implication orthoalgebra. For \/} $x,y\in A$
{\it define\/}
$$x\leq y \ {\it iff \/} \ x\bullet y=1$$
$$x\vee y:=(x\bullet y)\bullet y$$

\noindent
{\it and for\/} $p\in A$ {\it and\/} $a,b\in [p,1]$ {\it define\/}

\noindent
$$a\wedge b:=(((a\bullet p)\bullet (b\bullet p))\bullet (b\bullet p))\bullet p.$$

\noindent
{\it Then\/} $\leq $ {\it is an order on\/} $A$ {\it with the greatest element {\rm 1}, and\/}
$x\vee y=sup(x,y)$ {\it with respect to\/} $\leq ,$ {\it i.e. \/} $(A;\vee)$
{\it is a $\vee$-semilattice with the greatest element\/} 1. {\it For each\/} $p\in A$ {\it the
interval\/} $[p,1]$ {\it is a lattice with respect to\/} $\vee ,\wedge $ {\it as defined
above and\/} $a\bullet p$ {\it is an orthocomplement of\/} $a\in [p,1].$ {\it Hence,\/}
$(A;\vee )$ {\it is an orthosemilattice.\/}

\vskip8pt

Proof. By (a), $\leq $ is reflexive. Suppose $a\leq b$ and $b\leq a.$ Then $a\bullet b=1$
and $b\bullet a=1$ and we derive by (a) and (b) also
$a=1\bullet a=(b\bullet a)\bullet a=(a\bullet b)\bullet b=1\bullet b=b$,
thus $\leq $ is antisymmetric.

\noindent
Let $a\leq b$ and $b\leq c$. By (c) we have $1=b\bullet c\leq a\bullet c$ which yields
$a\bullet c=1,$ i.e. $a\leq c.$ Thus $\leq $ is also transitive, i.e. it is an order on $A.$

\noindent
Since $a\bullet 1=1$ by (a), 1 is the greatest element w.r.t. $\leq .$

\noindent
Put now $a\vee b:=(a\bullet b)\bullet b.$ If $a\leq b$ then $a\bullet b=1$ and hence
$a\vee b=(a\bullet b)\bullet b=1\bullet b=b,$ i.e.

\noindent
$(*)$ \hspace*{4cm} $a\leq b\Rightarrow a\vee b=b.$

\noindent
Further, $a\vee b=(a\bullet b)\bullet b\geq 1\bullet b=b$ (by (c)) and 
$a\vee b=b\vee a=(b\bullet a)\bullet a \geq 1\bullet a=a$ thus $a\leq a\vee b, b\leq a\vee b.$
Suppose $a\leq c, b\leq c.$ Then, by (c), $a\vee b=(a\bullet b)\bullet b\leq (c\bullet b)\bullet b=c\vee b.$
By $(*)$ and (b) we have $c\vee b=c,$ i.e. $a\vee b\leq c.$ We have shown that $a\vee b=sup(a,b)$ with
respect to $\leq.$
 
\noindent
Let $p\in A$ and $a,b\in [p,1].$ By (c) we obtain
$$a\leq b \ {\rm implies} \ b\bullet p\leq a\bullet p$$
$$a=a\vee p=(a\bullet p)\bullet p$$

\noindent
thus the mapping $a\longmapsto a\bullet p$ for $a\in [p,1]$ is an involutory antiautomorphism
of $([p,1],\leq)$ which implies De Morgan laws

\noindent
\hspace*{2cm} $(x\vee y)\bullet p=(x\bullet p)\wedge (y\bullet p) \ \ , \ \ (x\wedge  y)\bullet p=(x\bullet p)\vee (y\bullet p)$

\noindent
where $x\wedge  y:=((x\bullet p)\vee (y\bullet p))\bullet p.$

\noindent
This implies that $x\wedge y=inf(x,y)$ in $[p,1]$ w.r.t $\leq $ (restricted to the 
interval $[p,1]$). 

\noindent
Moreover, for $a\in [p,1]$ denote by $a^\bot _p$ the element $a\bullet p.$
Then $p\leq a$ implies
$$a=a\vee p=(a\bullet p)\bullet p=(a^\bot _p)^\bot _p$$
 
\noindent
and, by (d'),
$$a^\bot _p\vee a=((a\bullet p)\bullet a)\bullet a=1.$$

\noindent
Further,

\noindent
$a\wedge a^\bot _p=(a^\bot _p\vee (a^\bot _p)^\bot _p)^\bot _p=(a^\bot _p\vee a)^\bot _p=(a\vee a^\bot _p)^\bot _p=1^\bot _p=1\bullet p=p.$

\noindent
Hence, we have shown that $a^\bot _p$ is an orthocomplement of $a\in [p,1]$ in the
interval $[p,1].$
\hfill $\rule{2mm}{2mm}$

\vskip8pt

{\bf Corollary.} 
{\it Let\/} ${\cal A}=(A;\bullet ,1)$ {\it be an implication orthoalgebra. Then\/} ${\cal A}$
{\it is a set-theoretical union of strong ortholattices with the common greatest element\/} 1
{\it where the lattice operations coincide on the overlapping parts.\/}

\vskip8pt

In what follows, we give a certain description of congruences on implication orthoalgebras.
Consider a congruence $\Theta$ on an implication orthoalgebra 
${\cal A}=(A;\bullet ,1)$.
The class $[1]_\Theta$ will be called the {\bf kernel} of $\Theta$. Hence, each 
$\Theta \in Con({\cal A})$ determines its kernel. However, also vice versa, 
each congruence on $\cal A$ is uniquely determined by its kernel:

\vskip8pt

{\bf Theorem 3.} {\it Let ${\cal A}=(A;\bullet ,1)$ be an implication orthoalgebra and 
$\Theta, \Phi \in Con({\cal A})$. If $[1]_\Theta=[1]_\Phi$ then $\Theta =\Phi$.}

Proof. Assume $[1]_\Theta=[1]_\Phi$ for $\Theta, \Phi \in Con({\cal A})$ and let 
$(a,b)\in \Theta$. Then clearly

\noindent
\begin{center}
$\langle a\bullet b,1\rangle =\langle a\bullet b, a\bullet a\rangle \in \Theta$
\end{center}

\noindent
\begin{center}
$\langle b\bullet a,1\rangle =\langle b\bullet a, b\bullet b\rangle \in \Theta$
\end{center}

\noindent
thus $a\bullet b, b\bullet a\in [1]_\Theta=[1]_\Phi$ and hence 
 
\noindent
\begin{center}
$\langle a\bullet b,1\rangle \in \Phi$ and $\langle b\bullet a,1\rangle \in \Phi$. 
\end{center}

\noindent
Using Theorem 1(a),(b) we obtain 

\noindent
\begin{center}
$\langle (a\bullet b)\bullet b,b\rangle =\langle (a\bullet b)\bullet b, 1\bullet b\rangle\in \Phi,$
\end{center}

\noindent
\begin{center}
$\langle (b\bullet a)\bullet a,a\rangle =\langle (b\bullet a)\bullet a, 1\bullet a\rangle \in \Phi.$
\end{center}

\noindent
Hence
$$b\Phi (a\bullet b)\bullet b=(b\bullet a)\bullet a \Phi a$$
giving $(a,b)\in \Phi$, i.e. $\Theta \subseteq \Phi$. Analogously we can show 
$\Phi \subseteq \Theta$, thus $\Theta =\Phi.$
\hfill $\rule{2mm}{2mm}$

\vskip8pt

To describe a congruence $\Theta$ on an implication orthoalgebra $\cal A$, it is  enough 
to characterize its kernel $[1]_\Theta$. 

\vskip8pt

{\bf Theorem 4.} {\it Let ${\cal A}=(A;\bullet ,1)$ be an implication orthoalgebra and 
$D\subseteq A$ such that $1\in D$. The following conditions are equivalent:

\noindent
{\rm (1)} $D$ is a kernel of some $\Theta \in Con({\cal A})$;

\noindent
{\rm (2)} $D$ satisfies the following conditions:

\noindent
\hspace*{0.75cm}{\rm (D1)} if $x\in D$ and $y\bullet z\in D$ then $(x\bullet y)\bullet z\in D$

\noindent
\hspace*{0.75cm}{\rm (D2)} if $x\bullet y\in D$ and $y\bullet x\in D$ then 

\begin{center}
$(x\bullet z)\bullet (y\bullet z)\in D$ and $(z\bullet x)\bullet (z\bullet y)\in D$.
\end{center}}

Proof. It is an easy exercise to verify that every congruence kernel satisfies the 
conditions (D1) and (D2).

Conversely, let $1\in D\subseteq A$ and $D$ satisfy (D1) and (D2). Introduce a binary 
relation $\Theta_D$ on $A$ as follows:

\noindent
(A) \hspace*{3cm} $(x,y)\in \Theta_D \ \ {\rm iff} \ \ x\bullet y \ {\rm and} \ y\bullet x\in D.$

\noindent
Evidently, $\Theta_D$ is reflexive and symmetric. Suppose $(x,y)\in \Theta_D$ and 
$(y,z)\in \Theta_D$. Then $x\bullet y, y\bullet x, y\bullet z, z\bullet y \in D$ and, 
applying (c) of Theorem 1, we obtain

\noindent
(B) \hspace*{0.75cm} $((x\vee y)\bullet z)\bullet (x\bullet z)=(((x\bullet y)\bullet y)\bullet z)\bullet (x\bullet z)=1\in D.$

\noindent
Further, $x\bullet y\in D$ and $y\bullet z\in D$ imply by (D1) 

\noindent
(C) \hspace*{4cm} $((x\bullet y)\bullet y)\bullet z\in D.$

\noindent
Applying (D1) once more for $x=y$, we derive 
\begin{center}
$x\in D$ and $x\bullet z\in D$ imply $z=(x\bullet x)\bullet z\in D.$
\end{center}
We use the above rule together with (B) and (C) to obtain $x\bullet z\in D$. Analogously, 
one can show $z\bullet y\in D$, i.e. $(x,z)\in \Theta_D$ and $\Theta_D$ is also 
transitive. It is an easy calculation to show that (D2) together with the transitivity of $\Theta_D$ imply the substitution property 
with respect to $\bullet$, i.e. $\Theta_D$ is a congruence on $\cal A.$

It follows directly by (A) that $D$ is the kernel of $\Theta_D$.
\hfill $\rule{2mm}{2mm}$

\vskip8pt

In what follows, we are going to characterize congruence kernels as the so-called ideals.
Let ${\cal A}=(A;\bullet ,1)$ be an implication orthoalgebra. A subset $I\subseteq A$ is 
called an {\bf ideal} of $\cal A$ whenever there exists a congruence $\Theta$ on $\cal A$ such 
that $I$ is the kernel of $\Theta$. It is clear that each congruence $\Theta$ determines its 
kernel $[1]_\Theta$. However, also the converse statement is true by Theorem 3.

This result motivates us to describe ideals of implication orthoalgebras since every 
ideal determines just one congruence and every congruence is determined by an ideal.

For this, introduce the following concept adapted from [6]: a term 
\newline
$t(x_1,\dots,x_n,y_1,\dots,y_m)$ is called an {\bf ideal term} of ${\cal A}=(A;\bullet ,1)$ 
in $y_1,\dots,y_m$ whenever $t(x_1,\dots,x_n,1,\dots,1)=1$ is an identity in $\cal A$.

\vskip8pt

{\bf Lemma 1.} {\it Let $t(x_1,\dots,x_n,y_1,\dots,y_m)$ be an ideal term in $y_1,\dots,y_m$ of 
an implication orthoalgebra ${\cal A}=(A;\bullet ,1)$ and $I$ be an ideal of $\cal A$.
If $a_1,\dots,a_n\in A$ and $b_1,\dots,b_m\in I$ 
then $t(a_1,\dots,a_n,b_1,\dots,b_m)\in I$.}

Proof. Let $I$ be an ideal of $\cal A$. Then there exists a congruence $\Theta$ on $\cal A$ 
with $I=[1]_\Theta$. Assume further $a_1,\dots,a_n\in A$ and $b_1,\dots,b_m\in I$. Then
$\langle b_i,1\rangle \in \Theta$ for $i=1,\dots,m$ and hence
$$
\begin{array}{l}
\langle t(a_1,\dots,a_n,b_1,\dots,b_m),1\rangle =\\
\langle t(a_1,\dots,a_n,b_1,\dots,b_m),t(a_1,\dots,a_n,1,\dots,1)\rangle\in \Theta
\end{array}
$$
thus $t(a_1,\dots,a_n,b_1,\dots,b_m)\in [1]_\Theta =I$.
\hfill $\rule{2mm}{2mm}$

\vskip8pt

In other words, every ideal $I$ of $\cal A$ is {\bf closed under each ideal term} of $\cal A$.
Our goal is to show the crucial result, namely to prove that $I$ is an ideal of $\cal A$ 
iff $I$ is closed with respect to a {\bf finite} number of ideal terms which will be explicitly 
exhibited. Since every congruence kernel is closed with respect to substitutions (D1), (D2) 
as shown in Theorem 4, we need only to set up these terms and to verify that $I$ 
satisfies (D1), (D2) whenever it is closed with respect to them (the converse follows 
by Lemma 1).


{\bf Lemma 2.} {\it Let $I$ be a non-void subset of an implication orthoalgebra $\cal A$ 
closed under the following ideal terms of $\cal A$:

\noindent
$t_1(x,y)=x\bullet y$

\noindent
$t_2(x_1,x_2,y_1,y_2)=(x_1\bullet x_2)\bullet [y_2\bullet ((y_1\bullet x_1)\bullet x_2)]$

\noindent
$t_6(x,y_1,y_2)=(y_1\bullet (y_2\bullet x))\bullet x.$

\noindent
Then $I$ satisfies the implication {\rm (D1)}.}

Proof. At first we we show that $I$ satisfies the property

\noindent
(1) \hspace*{3.5cm} $a\in I$ and $a\bullet b\in I \ \Rightarrow \ b\in I$.

\noindent
Indeed, putting $y_1:=a\bullet b, y_2:=a, x:=b$ in the term $t_6$ we get
$$t_6(b,a\bullet b,a)=[(a\bullet b)\bullet (a\bullet b)]\bullet b=1\bullet b=b\in I$$
and (1) is proved.

Assume $x,y\bullet z\in I$. Since $x\in I$, the closedness of $I$ under $t_1$ gives us

\noindent
(2) \hspace*{2.5cm} $t_1(y\bullet x,x)=(y\bullet x)\bullet x=(x\bullet y)\bullet y\in I$.

\noindent
Analogously, taking $x_1:=y, x_2:=z, y_1:=x, y_2:=(x\bullet y)\bullet y$ in $t_2$ we obtain 

\noindent
(3) \hspace*{0.5cm} $t_2(y,z,x,(x\bullet y)\bullet y)=
    (y\bullet z)\bullet [((x\bullet y)\bullet y)\bullet ((x\bullet y)\bullet z)]\in I$.

\noindent
Further, $y\bullet z\in I$, hence applying (1) for $a:=y\bullet z$ and 
\newline
$b:=((x\bullet y)\bullet y)\bullet ((x\bullet y)\bullet z)$, we get
$$((x\bullet y)\bullet y)\bullet ((x\bullet y)\bullet z)\in I.$$
Finally, using (1) again for $a:=(x\bullet y)\bullet y$ and $b:=(x\bullet y)\bullet z$ gives 
us $(x\bullet y)\bullet z\in I$, finishing the proof. 
\hfill $\rule{2mm}{2mm}$ 

\vskip8pt

To guarantee the closedness of a given subset $I$ under the remaining property (D2), we 
need the following two lemmas:

\vskip8pt

{\bf Lemma 3.} {\it Let $I$ be a non-void subset of an implication orthoalgebra $\cal A$ 
closed under the ideal terms $t_6$ and 

\noindent
$t_3(x_1,x_2,y)=(x_1\bullet x_2)\bullet (x_1\bullet (y\bullet x_2))$;

\noindent
$t_4(x_1,x_2,x_3,y)=[(x_1\bullet x_2)\bullet (x_1\bullet (y\bullet x_3))]\bullet ((x_1\bullet x_2)\bullet (x_1\bullet x_3)).$

\noindent
Then $I$ has the property 
$$x\bullet y\in I \ {\rm and} \ y\bullet x\in I \ \Rightarrow \ (z\bullet x)\bullet (z\bullet y)\in I.$$}

Proof. Assume $x\bullet y, y\bullet x\in I$ for some $x,y\in A$. Using $t_3$ for 
$x_1:=z, x_2:=x, y:=y\bullet x$ we obtain

\noindent
(4) \hspace*{0.15cm} $t_3(z,x,y\bullet x)=(z\bullet x)\bullet [z\bullet ((y\bullet x)\bullet x)]\in I.$

Substituting $x_1:=z, x_2:=x, x_3:=y, y:=x\bullet y$ in $t_4$, we obtain

\noindent
(5) \hspace*{0.15cm} $t_4 (z,x,y,x\bullet y)=[(z\bullet x)\bullet (z\bullet ((x\bullet y)\bullet y)]\bullet ((z\bullet x)\bullet (z\bullet y))\in I.$

\noindent
The closedness of $I$ under $t_6$ guarantees by Lemma 2 that (1) holds for $I$, hence (4), (5) and (b) of Theorem 1 yield
$$(z\bullet x)\bullet (z\bullet y)\in I,$$
and we are done.
\hfill $\rule{2mm}{2mm}$ 

\vskip8pt

{\bf Lemma 4.} {\it Let $I$ be a non-void subset of an implication orthoalgebra $\cal A$ 
closed under the ideal terms $t_2,t_6$ and 

\noindent
$t_5(x_1,x_2,x_3,y)=[(x_1\bullet x_2)\bullet ((y\bullet x_3)\bullet x_2)]\bullet ((x_1\bullet x_2)\bullet (x_3\bullet x_2)).$

\noindent
Then $I$ has the property 
$$x\bullet y\in I \ {\rm and} \ y\bullet x\in I \ \Rightarrow \ (x\bullet z)\bullet (y\bullet z)\in I.$$}

Proof. The closedness of $I$ under $t_2$ immediately yields by putting $y_2:=1$ also the 
closedness under 
$$t'(x_1,x_2,y)=(x_1\bullet x_2)\bullet ((y\bullet x_1)\bullet x_2).$$
Let us substitute $x_1:=x, x_2:=z, y:=y\bullet x$ in $t'$. This gives us 
$$(x\bullet z)\bullet (((y\bullet x)\bullet x)\bullet z)\in I.$$
Moreover, $(y\bullet x)\bullet x=(x\bullet y)\bullet y$, hence also 

\noindent
(6) \hspace*{3.35cm} $(x\bullet z)\bullet (((x\bullet y)\bullet y)\bullet z)\in I.$

\noindent
Now, considering $t_5$ for instances $x_1:=x, x_2:=z, x_3:=y, y:=x\bullet y$, we have

\noindent
(7) \hspace*{1.7cm} $[(x\bullet z)\bullet (((x\bullet y)\bullet y)\bullet z)]\bullet ((x\bullet z)\bullet (y\bullet z))\in I$.

\noindent
The closedness of $I$ under $t_6$ gives us by Lemma 2 that $I$ satisfies the property (1). This together with 
(6) and (7) leads to 
$$(x\bullet z)\bullet (y\bullet z)\in I.$$
\hfill $\rule{2mm}{2mm}$ 

\vskip8pt

Applying the previous lemmas, we obtain the desired description of ideals in implication orthoalgebras:

\vskip8pt

{\bf Theorem 5.} {\it Let $I$ be a non-void subset of an implication orthoalgebra 
$\cal A$. Then $I$ is an ideal of $\cal A$ iff $I$ is closed with respect to the ideal 
terms $t_1, t_2, t_3, t_4, t_5, t_6.$}

\newpage

Authors' address:  Department of Algebra and Geometry\\
\hspace*{3.8cm}    Palack\'y University Olomouc\\
\hspace*{3.8cm}    Tomkova 40\\
\hspace*{3.8cm}    779 00 Olomouc\\
\hspace*{3.8cm}    Czech Republic\\

\vskip8pt

e-mail:            CHAJDA@RISC.UPOL.CZ\\
\hspace*{1.8cm}    HALAS@RISC.UPOL.CZ\\

\end{document}